\documentclass[conference]{IEEEtran}
\IEEEoverridecommandlockouts
\usepackage{cite}
\usepackage{amsmath,amssymb,amsfonts}
\usepackage{algorithmic}
\usepackage{graphicx}
\usepackage{textcomp}
\usepackage{xcolor}
\usepackage{graphicx}
\usepackage{amsmath}

\newcommand{\Rmnum}[1]{\expandafter\@slowromancap\romannumeral #1@}

\usepackage{fancyhdr} 
\usepackage{geometry}
\begin{document}
	
	\title{RIS-Assisted Received Adaptive Spatial Modulation for Wireless Communications\\
		\thanks{This work was supported by the Science and Technology Development Fund, Macau, SAR, under Grant 0044/2022/A1. \textsuperscript{*}Corresponding author: Benjamin K. Ng (bng@mpu.edu.mo).}
	}
	
	\author{  
		\IEEEauthorblockN{Chaorong Zhang\IEEEauthorrefmark{2}\textsuperscript{1},  
			Hui Xu\IEEEauthorrefmark{2}\textsuperscript{2},  
			Benjamin K. Ng\IEEEauthorrefmark{2}\textsuperscript{3*},  
			Chan-Tong Lam\IEEEauthorrefmark{2}\textsuperscript{4},
			and Ke Wang\IEEEauthorrefmark{2}\textsuperscript{5}}  
		\IEEEauthorblockA{\textit{\IEEEauthorrefmark{2}Faculty of Applied Sciences, Macao Polytechnic University, Macao SAR, China}
		\\
			\textsuperscript{1}p2314785@mpu.edu.mo,  
			\textsuperscript{2}p2112282@mpu.edu.mo,  
			\textsuperscript{3}bng@mpu.edu.mo,  
			\textsuperscript{4}ctlam@mpu.edu.mo,
			\textsuperscript{5}kewang@mpu.edu.mo}  
	}

	\maketitle
	
	\begin{abstract}
		A novel wireless transmission scheme, as named the reconfigurable intelligent surface (RIS)-assisted received adaptive spatial modulation (RASM) scheme, is proposed in this paper.
		In this scheme, the adaptive spatial modulation (ASM)-based antennas selection  works at the receiver by employing the characteristics of the RIS in each time slot, where the signal-to-noise ratio at specific selected antennas can be further enhanced with near few powers. 
		Besides for the bits from constellation symbols, the extra bits can be mapped into the indices of receive antenna combinations and conveyed to the receiver through the ASM-based antenna-combination selection, thus providing higher spectral efficiency.
		To explicitly present the RASM scheme, the analytical performance of bit error rate of it is discussed in this paper. 
		As a trade-off selection, the proposed scheme shows higher spectral efficiency and remains the satisfactory error performance.
		Simulation and analytical results demonstrate the better performance and exhibit more potential to apply in practical wireless communication.
	\end{abstract}
	
	
	\begin{figure*}
		\centering
		\includegraphics[width=17cm,height=5.3cm]{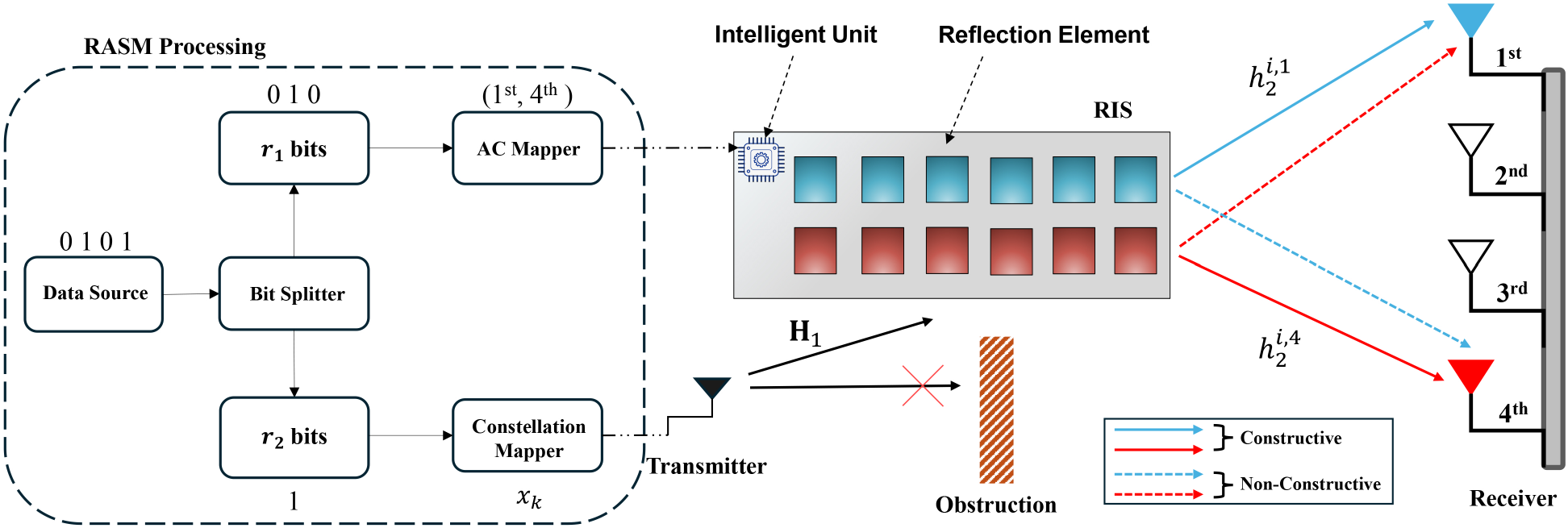}\\
		\caption{System model for the RASM scheme.}
	\end{figure*}
	
	\vspace{-7pt}
	\section{Introduction}
	A popular technique with the potential to revolutionize conventional wireless communication paradigm, called the index modulation (IM), has plenty of branches, such as, spatial modulation (SM), space shift keying (SSK) schemes, and so on, which are the traditional ones with the indices mapped into the transmit antennas \cite{ref1}. 
	Later, the generalized spatial modulation (GSM) and generalized space shift keying (GSSK) schemes are proposed to address the antenna expenses problem, but where the transmit antennas are activated only in a fix number and the bit-error rate (BER) performance deteriorates significantly compared with the SM and SSK scheme. 
	Hence, a groundbreaking IM approach, referred to as adaptive spatial modulation (ASM), is put forward to tackle the issues \cite{ref2}-\cite{ref3}. 
	Within the ASM scheme, the activation in the number of transmit antennas can be flexibly adjusted, which can be futher enhance higher spectral efficiency (SE) compared to the other IM schemes.
	As a trade-off selection in wireless communication, the ASM scheme provides better error performance than those traditional IM schemes, but brings some issues, e.g., sacrificing robustness in resisting the inter-channel interference (ICI) and the inter-antenna synchronization (IAS), increasing number of RF chains at transmitter, etc., as a few of antennas are likely to be activated simultaneously. 
	To solve these issues, a novel technique is considered to be applied in the ASM scheme to supplement and  improve the system design.
	\par
	As a promising technique, the reconfigurable intelligent surface (RIS) incorporates a chip, known as the intelligent unit, 
	which can control the angles of reflection elements (REs) to adjust the phase shift of transmitted signals with near few powers, i.e. amplifying the transmitted power at receiver [4], where the signal-to-noise ratio (SNR) can be further enlarged at receiver.
	By possessing these advantages, the RIS can achieve the antenna selection at the receive antennas by adjusting the phase to amplify the power at one or more specific antennas, which can more effectively address the issues of the ASM scheme mentioned above \cite{ref5}-\cite{ref10}. 
	Capable of working not only at the transmit antenna but also at the receiver through the RIS \cite{ref9}-\cite{ref11}, which is called the received IM scheme in this paper, the antennas selection working at the receive antennas can also convey the bits mapped by the antenna indices.
	Compared with the conventional transmitted IM schemes, the received IM schemes present promising avenues for advancing both SE and  BER performance beyond conventional paradigms without the RIS.
	Also, it can use less number of transmit antennas compared to some RIS-aided transmitted IM schemes, e.g., transmitted ASM (TASM) scheme, transmitted GSM scheme, and so on.
	In \cite{ref7}, by assisting with the RIS, the received SM (RSM) scheme is first proposed with antenna selection working at receive antennas, where the BER performance is satisfactory but the wireless channel from the transmitter to the RIS is not considered and the SE still can be further improved. 
	Later, the RIS-assisted various SM (RGSM) and GSSK (RGSSK) schemes are proposed in \cite{ref8}-\cite{ref10} to further enhance the SE with proposing two novel low-complexity detectors respectively, where the BER performance of the proposed schemes outperform that of the transmitted IM ones,  but the BER performance seems worse than that of the RSM and RSSK schemes as well the number of selecting antennas fixing. 
	Thus, we proposed the RIS-assisted received adaptive spatial modulation (RASM) schemes to address above issues.
	\par
	In light of these state-of-arts, the main contributions of this work can be briefly summarized as follows: 
	1) A novel RIS-assisted IM scheme with random antenna combination (AC) selection is first proposed in this paper, referred to as the RASM scheme, aiming at improving the performance of SE and BER, as well solving the problems of ICI and IAS in the transmitted IM schemes, etc.
	2)  The theoretical analysis for error performance in the RASM scheme with the random AC selection and maximum likelihood (ML) detector is provided, of which some valuable insights and interesting findings are also given. 
	3) Results from the Monte Carlo simulations are presented to demonstrate some interesting findings and elaborate the benefits of the RASM scheme. 
	
	\begin{table}[!] 
		\centering 
		\caption{Mapping table for selected ACs}
		\resizebox{.8\columnwidth}{!}{ 
			\begin{tabular}{|c|c|c|c|} \hline 
				Bits & Antennas & Bits & Antennas   \\ \hline
				0 0 0 & (1st) & 1 0 0 & (1st,2nd,3rd)  \\ \hline
				0 0 1 & (2nd,3rd) &	1 0 1 &	(1st,2nd,4st) \\ \hline
				0 1 0 & (1st,4th) & 1 1 0 & (1st,3rd,4th) \\ \hline
				0 1 1 & (3rd,4th) & 1 1 1 & (2nd,3rd,4th) \\ \hline
			\end{tabular}
		}
	\end{table}
	
	\vspace{-7pt}
	\section{System Model}
	
	\vspace{-7pt}
	\subsection{Configurations and Wireless Channels}
	System model of the proposed scheme is illustrated in this section, as shown in Fig. 1, where the single-input multiple-output (SIMO) communication is considered. 
	The transmitter equips one transmit antenna and the receiver has $N_r$ receive antennas. 
	Also, a shadow area is considered. Specifically, the direct link between the transmitter and the receiver is assumed to be blocked by a huge obstacle, thus the only way that can connect them is the link through the RIS. The RIS has $N$ REs and one intelligent unit connected to the transmitter. 
	As one of the examples in the Table \(\text{I}\) and Fig. 1, REs can amplify the signals at the first and fourth receive antenna through the passive beamforming, of which the first and fourth antennas can be considered as a AC and selected as the index to represent the bits (010).
	Here, the total number of antennas in each AC can be represented as $N_a$ with $1\le N_a\le N_r$, of which we define each AC as $V_r$ with total $N_a$ selected receive antennas, as $V_r=\left\{ 1,3,...,l \right\}$ where $l$ represents $l$-th selected antenna.
	The full set of ACs is given as $\varLambda (r)$, with $r\in \left\{ 1,2,\cdots ,J \right\}$ representing $r$-th selected AC and $J$ representing the total number of ACs. Thus, $J$ is expressed as
	\begin{align}
		J=\sum_{N_a=1}^{N_r}{\left( \begin{array}{c}	N_r\\	N_a\\\end{array} \right)}=2^{N_r}-1.
	\end{align}
	By considering the $M$-ary constellation symbols, $b$ bit per channel use (bpcu) are needed in the RASM, where $b=b_1+b_2$ with $b_1=\mathrm{floor}\left( \log _2\!\:J \right)$ and $b_2=\log _2\!\:M$.
	\par
	In this system model, by assuming the far-field wireless communications in city environment, the channel matrix channel for the channel state information (CSI) of the channel from the transmitter to the RIS is defined as $\mathbf{H}_1\in \mathbb{C} ^{N\times 1}$, another one from the RIS to the receiver is represented as $\mathbf{H}_2\in \mathbb{C} ^{N_r\times N}$, which both follow the Rayleigh flat fading with zero mean value and unit variance. 
	The diagonal matrix of adjusted phase in the RIS can be given as $\mathbf{\Phi }=\mathsf{diag}(\exp \!\:\left( j\varphi _{1,m} \right) ,\exp \!\:\left( j\varphi _{2,m} \right) ,\cdots ,\exp \!\:\left( j\varphi _{N,m} \right) )$, with $\varphi _{i,m}$ representing the adjusted phase shift of $i$-th reflecting element and for $m$-th receive antenna.
	Furthermore, we define the wireless channel from the transmit antenna to the $i$-th reflecting element and the one from the $i$-th reflecting to the $m$-th receive antenna as $h_{1}^{i}$ and $h_{2}^{i,m}$ with $i=1,2,\cdots ,\mathrm{ }N$ and $m=1,2,\cdots ,\mathrm{ }N_r$, respectively. 
	$h_{1}^{i}$ and $h_{2}^{i,m}$ are assumed to be independent and identically distributed complex Gaussian random variables with $\mathcal{C} \mathcal{N} (0,1)$, which can be extended to $h_{1}^{i}=\alpha _ie^{-j\theta _i}$ and $h_{2}^{i,m}=\beta _{i,m}e^{-j\omega _{i,m}}$, with $\alpha $ and $\beta $ representing the Rayleigh factors in various channels, as well $\theta $ and $\omega $ representing the phase shift of these channels.
	\par
	Based on the principle of ASM scheme, although $N_r$ receive antennas are equipped at receiver, only $N_s$ antennas can be selected to convey transmitted signals. 
	The REs are evenly divided into $N_s$ parts, which has $N_E=\lfloor \frac{N}{N_s} \rfloor $ number of REs, and each part is corresponding for one selected receive antenna to achieve the passive beamforming. 
	Besides, each antenna not only can receive the signals reflected from the REs that are corresponding for itself but also from the non-corresponding ones. 
	Specifically, as shown in Fig.1, the first receive antenna can receive the signals reflected from the REs corresponding for itself (as shown in the blue dashed line) and can be affected by those for the fourth receive antenna (as shown in red dashed line) in the meanwhile,
	which leads to the constructive part and non-constructive part in the signals for $m$-th receive antenna, respectively. 
	
	\vspace{-7pt}
	\subsection{Random AC Selection}
	Since the available ACs are more than the required and must be a power of 2, e.g., 3 receive antennas lead to 7 ACs but 4 ones are available to be represented the transmitted bit sequences, an algorithm named as the random AC selection is design in the proposed scheme. 
	Specifically, before the transmission, the transmitter and receiver have an agreement that only $D$ ACs can be the indices to convey the bits sequence, where $D=\log _2\!\:b_1$ and $D$ ACs are selected by the designed random AC selection. 
	For example, we select 8 ACs randomly from the 15 available ACs, without any specific order or regulation. 
	After the random AC selection, the AC set can be given as $\varLambda (r)$ with $r=1,\mathrm{ }2,\cdots ,D$. 
	The random AC selection works at the intelligent unit of the RIS after the information of selected ACs inputting. 
	Then, the intelligent unit controls various part/parts of REs to achieve the passive beamforming for the specific receive antenna/antennas, respectively.
	
	\vspace{-7pt}
	\subsection{Transmission} 
	For the selected AC, each one represents different bit sequence, e.g., the third selected AC indicates the bits 010.
	Moreover, the last bit in sequence is 1, which is represented by $k$-th constellation symbol $x_k$ in $M$-ary PSK/QAM constellation symbol vector as $\boldsymbol{x}_k=\left[ x_1,x_2,\cdots ,x_M \right] ^T$ as shown in Fig. 1. Thus, the expression of received signal in RASM can be obtained as
	\begin{align}
		\boldsymbol{y}=\mathbf{H}_2\mathbf{\Phi }_r\mathbf{H}_1x_k+\mathbf{N},
	\end{align}
	where $\mathbf{\Phi }_r=diag\{\Phi _{1,l_1},\Phi _{2,l_1},\cdots ,\Phi _{z,l_{N_a}}\}\in \mathbb{C} ^{N\times N}$ represents the diagonal matrix of adjusted phase for $r$-th AC $\footnote{To make sure that the receive antennas in selected AC have the equal amplification from the RIS in each time slot, the configuration of adjusted phase from $z$-th part of REs to receiver is given as $\mathbf{\Phi } _{z,l}=\exp \left( -\jmath\mathrm{phase}\left( \mathbf{H}_2\left[ z,\left( z-1 \right) \Delta +:z\Delta \right] \mathbf{H}_1 \right) \right)$ according to [18], where $\Delta =\lfloor \frac{N}{N_a} \rfloor$, and $z$-th part of REs is corresponding for the specific selected AC with $\left[ \left( z-1 \right) N_E+1:zN_E \right] $ number of elements. This is out of scope, thus it is not extended in this paper.}$
	and the $\mathbf{N}\in \mathbb{C} ^{N_r\times 1}$ is the matrix of Gaussian white noise. 
	However, to better show the differences between various indices of ACs, we give the received signal expression at $l$-th receive antenna in $r$-th selected AC without Gaussian white noise $\tau_l$, which is expressed as 
	$\footnote{We assume that each part of REs can perfectly achieve the passive beamforming for specific $l$-th selected antenna as $\varphi _i=\omega _{i,l}+\theta _i$ [9-12].}$ 
	\begin{small}
		\begin{align}
			&y'_l=\underset{\textit{constructive\,\,part}}{\underbrace{\sum_{i=\left( l-1 \right) N_E+1}^{zN_E}{\mathrm{h}_{2}^{i,l}\Phi _{i,l}\mathrm{h}_{1}^{i}}x_k}}+\underset{\textit{non-constructive\,\,part}}{\underbrace{\sum_{q=1,\mathrm{ }q\ne l}^{N_a}{\sum_{i=\left( q-1 \right) N_E+1}^{qN_E}{\mathrm{h}_{2}^{i,l}\Phi _{i,l}\mathrm{h}_{1}^{i}}x_k}}}  \nonumber
			\\
			&=\underset{\textit{non-constructive\,\,part}}{\underbrace{\sum_{q=1,\mathrm{ }q\ne z}^{N_a}{\sum_{i=\left( q-1 \right) N_E+1}^{qN_E}{\beta _{i,l}\alpha _ie^{-j\varPsi _{i,l}}x_k}}}}+\underset{\textit{constructive\,\,part}}{\underbrace{\sum_{i=\left( l-1 \right) N_E+1}^{lN_E}{\beta _{i,l}\alpha _ix_k}}},
		\end{align}
	\end{small}where $\Phi _{i,l}$ can be simplified as $\Phi _{i,l}=\exp \!\:\left( j\varphi _{i,l} \right) $ representing the adjust phase of the $i$-th RE and for the $l$-th selected receive antenna in selected AC, $\varPsi_{i,l}=\varphi _i-\omega _{i,l}+\theta _i$, $\mathrm{ }l\in \left\{ 1,\mathrm{ }\cdots ,\mathrm{ }N_a \right\} $,
	and $n_l$ is additive white Gaussian noise at $l$-th selected receive antenna in selected AC with $\mathcal{C} \mathcal{N} \left( 0,N_0 \right) $ distribution.
	In the meanwhile, the complete received signal at $l$-th selected antenna is given as $y_l=y'_l+\tau _l$ with $\tau_l$ representing the Gaussian white noise at $l$ AC.
	And the received signal of $u$-th unselected received antenna can be given as
	\begin{align}
		y_u=\sum_{q=1}^{N_a}{\sum_{i=\left( q-1 \right) N_E+1}^{qN_E}{\beta _{i,u}\alpha _ie^{-j\left( \varphi _i-\omega _{i,l}-\theta _i \right)}x_k}}+\tau_u,
	\end{align}
	with which the SNR can be obtained as
	\begin{align}
		\gamma _u=\frac{\left\| \sum_{q=1}^{N_a}{\sum_{i=\left( q-1 \right) N_E+1}^{qN_E}{\beta _{i,u}\alpha _ie^{-j\left( \varphi _i-\omega _{i,l}-\theta _i \right)}x_k}} \right\| ^2}{N_0}.
	\end{align}
	According to the characteristic of the RIS, when $\varphi _i=\omega _{i,l}+\theta _i$, the SNR at $l$-th selected antenna can be maximized, which is given as
	\begin{footnotesize}
		\begin{align}
			\left\{ \gamma _l \right\} _{max}&=\frac{\left| \sum_{i=\left( l-1 \right) N_E+1}^{lN_E}{\beta _{i,l}\alpha _ix_k} \right|^2}{N_0} \nonumber
			\\
			&+\frac{\left| \sum_{q=1,\mathrm{ }q\ne l}^{N_a}{\sum_{i=\left( q-1 \right) N_E+1}^{qN_E}{\beta _{i,l}\alpha _ie^{-j\left( \varphi _i-\omega _{i,l}-\theta _i \right)}x_k}} \right|^2}{N_0}.
		\end{align}
	\end{footnotesize}The ML detector of the RASM scheme can be given as:
	\begin{align}
	\left\{ \hat{r} \right\} =\mathrm{arg}\min_{r} \!\:\left\| \boldsymbol{Y}-\mathbf{G}_{r} \right\| _{2}^{2}
	\end{align}
	with $\mathbf{G}_{r} = E_s\mathbf{H}_2\mathbf{\Phi }_r\mathbf{H}_1$.
	Meanwhile, $\mathbf{G}_{r}$ can be rewritten as $\mathbf{G}_{r}=\left\{ G_n \right\} _{n=1}^{N_r}$, where $G_n$ can be expressed as
	\begin{align}
	G_n=\begin{cases}
		E_s\left\{ \sum_{i=\left( l-1 \right) N_E+1}^{lN_E}{\mathrm{h}_{2}^{i,n}\mathrm{h}_{1}^{i}}+G_{n}^{\prime} \right\} , v_{r,n}=1\\
		\\
		E_s\left\{ \sum_{q=1}^{N_a}{\sum_{i=\left( q-1 \right) N_E+1}^{qN_E}{\beta _{i,u}\alpha _ie^{\jmath\varPsi _{i,l}}}} \right\} , v_{r,n}=0\\
	\end{cases}
	\end{align}
	with $G_{n}^{\prime}=E_s\sum_{q=1,q\ne l}^{N_a}{\sum_{i=\varsigma}^{qN_E}{\mathrm{h}_{2}^{i,n}\Phi _{i,l}\mathrm{h}_{1}^{i}}}$.
	
	\vspace{-7pt}
\section{Performance Analysis}
This section provides the analytical performance in average BER (ABER) of RASM schemes. To better present the theoretical ABER with the ML detector, we assume the channel fading is the Rayleigh fading and the receiver has the perfect knowledge of CSI. 
According to Eq. (7) of ML detector, the instantaneous pairwise error probability (PEP) can be expressed by giving channel $\mathrm{h}_{1}^{i}$ and $\mathrm{h}_{2}^{i,l}$ as
\begin{align}
	&\mathrm{P_r}\left( \left\{ r,k \right\} \rightarrow \{\hat{r},\hat{k}\}\mid \mathrm{h}_{1}^{i},\mathrm{h}_{2}^{i,m} \right)  \nonumber
	\\
	&=\mathrm{Pr}\!\:\left( \sum_{n=1}^{N_r}{\left\| y_n-G_n \right\| _{2}^{2}}>\sum_{n=1}^{N_r}{\left\| y_n-\hat{G}_n \right\| _{2}^{2}} \right), 
\end{align}
where $\hat{G}_n$ represents the estimated $G_n$ and $y_n$ represents received signal at $n$-th receive antenna.
After some algebraic operations, Eq. (9) can be extended and rewritten as
\begin{align}
	&\mathrm{P_r}\left( \left\{ r,k \right\} \rightarrow \{\hat{r},\hat{k}\}\mid \mathrm{h}_{1}^{i},\mathrm{h}_{2}^{i,m} \right)  =\mathrm{Pr}\!\:\left( \Gamma <0 \right) ,
\end{align}
with
\begin{align}
	\Gamma =\sum_{n=1}^{N_r}{\left\| G_n-\hat{G}_n \right\| _{2}^{2}}+\sum_{n=1}^{N_r}{2\mathfrak{N} \left\{ {\tau _n}^*\left[ G_n-\hat{G}_n \right] \right\}},
\end{align}
where ${\tau_l}^*$ represents the complex conjugation of Gaussian white noise and $\Gamma \sim \mathcal{N} \left( \mu _{\Gamma},\sigma _{\Gamma}^{2} \right) $ 
with $Z=\sum_{n=1}^{N_r}{\left\| G_n-\hat{G}_n \right\| _{2}^{2}}$
and $\sigma _{\Gamma}^{2}=2N_0\sum_{n=1}^{N_r}{\left\| G_n-\hat{G}_n \right\| _{2}^{2}}$. 
Thus, by applying the Gaussian Q-function as $Q(\frac{-\mu _{\Gamma}}{\sigma _{\Gamma}^{2}})$ the PEP can be further denoted as follows:
\begin{align}
	\mathrm{P_r}\left( \left\{ r,k \right\} \rightarrow \{\hat{r},\hat{k}\}\mid \mathrm{h}_{}^{i},\mathrm{h}_{2}^{i,m} \right) 
	=Q\left( \sqrt{\frac{Z}{2N_0}} \right).
\end{align} 
Considering the alternative form of Q-function, the unconditional PEP, which is averaged over channel coefficients as $\mathrm{P_r}\left( \left\{ r,k \right\} \rightarrow \{\hat{r},\hat{k}\} \right) =\mathbb{E}_Z\left[ Q\left( \sqrt{\frac{Z}{2N_0}} \right) \right] $, can be further calculated as
\begin{align}
	\mathrm{P_r}\left( \left\{ r,k \right\} \rightarrow \{\hat{r},\hat{k}\} \right) &=\int_0^{\infty}{Q\left( \sqrt{\frac{Z}{2N_0}} \right) f_Z\left( Z \right) dZ} \nonumber
	\\
	&=\frac{1}{\pi}\int_0^{\frac{\pi}{2}}{M_Z\left( \frac{-1}{4\sin ^2\!\:\tau N_0} \right) d\tau},
\end{align}
which applies the moment generating function (MGF) of $Z$, given as $M_Z\left( t \right) =\int_0^{\infty}{e^{tZ}dZ}$, with $t=\frac{-1}{4\sin ^2\!\:\tau N_0}$. 
Actually, $Z$ is the squared Euclidean distance (SED) between the symbols $G_{r,k}$ and $G_{\hat{r},\hat{k}}$. 
Here, the MGF of Eq. (13) can be derived by considering the general quadratic form of correlated Gaussian random variables and counts on erroneous or correct detection of the $r$-th and $\hat{r}$-th AC indices. 
Note that the error in the detection of AC index is able to affect the accuracy of constellation symbol detection. Thus, we separate $Z$ into two categories, which are given as: $Z_1=\left\{ Z \mid r\ne \hat{r} \right\} $ and $Z_2=\left\{ Z \mid r=\hat{r} \right\} $.
\par
\noindent 1) First case $Z_1$: $\left\{ Z \mid r\ne \hat{r} \right\} $
\par
In this case, the SED can be given as $Z_1=Z_{1}^{1}+Z_{1}^{2}+Z_{1}^{3}$, which are given specifically in Eq. (14), (15), and (16) with $\varphi _i-\omega _{i,l}-\theta _i=\varPsi_{i,l} $, as shown at the top of next page.
\begin{figure*}[!] 
	\centering 
	\vspace*{0pt} 
	\begin{align}
		Z_{1}^{1}=\sum\nolimits_{l\in V_r}^{}{\left| \left( \sum_{i=\left( l-1 \right) N_E+1}^{lN_E}{\beta _{i,l}\alpha _i}+\sum_{q=1,\mathrm{ }q\ne l}^{N_a}{\sum_{i=\left( q-1 \right) N_E+1}^{qN_E}{\beta _{i,l}\alpha _ie^{\jmath\varPsi _{i,l}}}} \right) x_k-\sum_{q=1}^{N_a}{\sum_{i=\left( q-1 \right) N_E+1}^{qN_E}{\beta _{i,l}\alpha _ie^{\jmath\varPsi _{i,l}}x_{\hat{k}}}} \right|}^2,
	\end{align}
	
	\begin{align}
		Z_{1}^{2}=\sum\nolimits_{{\hat{l}}\in V_{\hat{r}}}^{}{\left| \sum_{q=1}^{N_a}{\sum_{i=\left( q-1 \right) N_E+1}^{qN_E}{\beta _{i,\hat{l}}\alpha _ie^{\jmath\varPsi _{i,\hat{l}}}x_k}}-\left( \sum_{i=\left( l-1 \right) N_E+1}^{lN_E}{\beta _{i,\hat{l}}\alpha _i}+\sum_{q=1,\mathrm{ }q\ne l}^{N_a}{\sum_{i=\left( q-1 \right) N_E+1}^{qN_E}{\beta _{i,\hat{l}}\alpha _ie^{\jmath\varPsi _{i,\hat{l}}}}} \right) x_{\hat{k}} \right|}^2,
	\end{align}
	
	\begin{align}
		Z_{1}^{3}=\sum_{n=1,n\ne l,n\ne \hat{l}}^{N_r}{\left| \sum_{q=1}^{N_a}{\sum_{i=\left( q-1 \right) N_E+1}^{qN_E}{\left( \beta _{i,n}\alpha _ie^{\jmath\left( \varphi _i-\omega _{i,l}-\theta _i \right)}x_k-\beta _{i,n}\alpha _ie^{\jmath\left( \varphi _i-\omega _{i,l}-\theta _i \right)}x_{\hat{k}} \right)}} \right|^2}.
	\end{align}
	\hrulefill 
\end{figure*}

Here, $Z_{1}^{1}$, $Z_{1}^{2}$, and $Z_{1}^{3}$ stand for the $m=r$, $m=\hat{r}$, and $m\ne r,\mathrm{ }\hat{r}$ respectively. 
Then, we can denote $Z_{1}^{1}$ and $Z_{1}^{2}$ as
\begin{align}
	Z_{1}^{1}=\left| q_1 \right|^2=\left( q_1 \right) _{\mathfrak{N}}^{2}+\left( q_1 \right) _{\mathfrak{T}}^{2},
\end{align}
\begin{align}
	Z_{1}^{2}=\left| q_2 \right|^2=\left( q_2 \right) _{\mathfrak{N}}^{2}+\left( q_2 \right) _{\mathfrak{T}}^{2},
\end{align}
where $q_1$ and $q_2$ follow complex Gaussian distribution by applying central limit theorem (CLT). 
Note that $\alpha $ and $\beta $ are the magnitudes of standard complex Gaussian random variables, they follow the Rayleigh distribution with mean value $\frac{\sqrt{\pi}}{2}$ variance $\frac{4-\pi}{4}$, respectively. 
Moreover, since $\omega _i$, $\varphi _i$, and $\theta _i$ are the phase of standard complex Gaussian random variable and follow the uniform distribution at the range in $(-\pi ,\pi )$, 
the mean value and the variance of these ones are given as 0 and 0.5 respectively, which means $\mathbb{E}\left[ \beta _{i,m}\alpha _ie^{-\jmath\varPsi_{i,l})} \right] =0$ 
and $\mathbb{D}\left[ \beta _{i,m}\alpha _ie^{-\jmath\varPsi_{i,l}} \right] =1$. 
Without loss of generality, the number of antennas in the estimated AC is equal to the ones in the selected one. 
Therefore, the mean vector $\boldsymbol{\mu }_1$ and covariance matrix $\boldsymbol{\sigma }_{1}^{2}$ of $\boldsymbol{z}_1=\left[ \left( q_1 \right) _{\mathfrak{R}},\left( q_1 \right) _{\mathfrak{T}},\left( q_2 \right) _{\mathfrak{R}},\left( q_2 \right) _{\mathfrak{T}}\mathrm{ } \right] $, which are respectively given as follows:
\begin{align}
	\boldsymbol{\mu }_1=\frac{N\pi}{4}\left[ \left( x_k \right) _{\mathfrak{N}},\left( x_k \right) _{\mathfrak{T}},-\left( x_k \right) _{\mathfrak{N}},-\left( x_k \right) _{\mathfrak{T}} \right] ^\mathsf{T},
\end{align}
\begin{align}
	\boldsymbol{\sigma }_{1}^{2}=
	\begin{bmatrix}
		\vspace{1ex}
		\sigma _{1}^{2}&	\quad	\sigma _{1,2}^{2}&	\quad	\sigma _{1,3}^{2}&	\quad	\sigma _{1,4}^{2}\\
		\vspace{1ex}
		\sigma _{1,2}^{2}&	\quad	\sigma _{2}^{2}&	\quad	\sigma _{2,3}^{2}&	\quad	\sigma _{2,4}^{2}\\
		\vspace{1ex}
		\sigma _{1,3}^{2}&	\quad	\sigma _{2,3}^{2}&	\quad	\sigma _{3}^{2}&	\quad	\sigma _{3,4}^{2}\\
		\vspace{1ex}
		\sigma _{1,4}^{2}&	\quad	\sigma _{2,4}^{2}&	\quad	\sigma _{3,4}^{2}&	\quad	\sigma _{4}^{2}\\
	\end{bmatrix} ,
\end{align}
where 
{\small
	\begin{align}
		\sigma _{1}^{2}=N_a\left[ \left( N_a-\frac{\pi ^2}{16} \right) N_E\left( x_k \right) _{\mathfrak{R}}^{2}+\frac{N\left| x_{\hat{k}} \right|^2\mathrm{ }}{2} \right] , \nonumber
	\end{align}
	\begin{align}
		\sigma _{2}^{2}=N_a\left[ \left( N_a-\frac{\pi ^2}{16} \right) N_E\left( x_k \right) _{\mathfrak{T}}^{2}+\frac{N\left| x_{\hat{k}} \right|^2\mathrm{ }}{2} \right] , \nonumber
	\end{align}
	\begin{align}
		\sigma _{3}^{2}=N_a\left[ \left( N_a-\frac{\pi ^2}{16} \right) N_E\left( x_{\hat{k}} \right) _{\mathfrak{R}}^{2}+\frac{N\left| x_k \right|^2\mathrm{ }}{2} \right] , \nonumber
	\end{align}
	\begin{align}
		\sigma _{4}^{2}=N_a\left[ \left( N_a-\frac{\pi ^2}{16} \right) N_E\left( x_{\hat{k}} \right) _{\mathfrak{T}}^{2}+\frac{N\left| x_k \right|^2\mathrm{ }}{2} \right] .
	\end{align}
}By applying the property of covariance between two random variables $X$ and $Y$ as
\begin{align}
	2\mathrm{Cov}\left( X,Y \right) =\mathbb{D}\left[ X+Y \right] -\mathbb{D}\left[ X \right] -\mathbb{D}\left[ Y \right] ,
\end{align}
we can easily obtain the $\sigma _{1,2}^{2}$, $\sigma _{3,4}^{2}$, $\sigma _{1,4}^{2}$, $\sigma _{2,3}^{2}$, and $\sigma _{2,4}^{2}$.
The MGF of the generalized non-central chi-square distribution is given as follows \cite{ref11} and \cite{ref7}:
\begin{align}
	&M_X\left( t \mid \boldsymbol{\mu },\boldsymbol{\sigma }^2 \right) =\left[ \det \left( \mathbf{E}-2t\boldsymbol{\sigma }^2 \right) \right] ^{-\frac{1}{2}}\times \nonumber
	\\
	&\exp \left\{ -\frac{1}{2}\boldsymbol{\mu }^\mathsf{T}\left[ \mathbf{E}-\left( \mathbf{E}-2t\boldsymbol{\sigma }^2 \right) ^{-1} \right] {\boldsymbol{\sigma }}^{-1}\boldsymbol{\mu } \right\} ,
\end{align}
where $X=\sum_{f=1}^f{X_{f}^{2}}$, is the unit matrix, $\boldsymbol{\mu }$ and $\boldsymbol{\sigma }^2$ represent the mean vector and covariance matrix of $\left[ X_1,X_2,\cdots ,X_f \right] ^\mathsf{T}$. Then, substituting Eq. (19) and (20) into Eq. (23), we can yield the MGF of $Z_{1}^{1}+Z_{1}^{2}$, given as $M_{Z_{1}^{1}+Z_{1}^{2}}\left( t \mid \boldsymbol{\mu }_1,\boldsymbol{\sigma }_{1}^{2} \right) $.
Nevertheless, the MGF of $Z_{1}^{3}$ is still needed to be derived. In Eq. (16), we define that
\begin{align}
	\Gamma =&\sum_{q=1}^{N_a}{\sum_{i=\left( q-1 \right) N_E+1}^{qN_E}{\beta _{i,n}\alpha _ie^{\jmath\left( \varphi _i-\omega _{i,l}-\theta _i \right)}x_k}}  \nonumber
	\\
	&-\sum_{q=1}^{N_a}{\sum_{i=\left( q-1 \right) N_E+1}^{qN_E}{\beta _{i,n}\alpha _ie^{\jmath\left( \varphi _i-\omega _{i,l}-\theta _i \right)}x_{\hat{k}}}},
\end{align}
which has variance $\sigma _{\Gamma}^{2}=\frac{N\left( \left| x_k \right|^2+\left| x_{\hat{k}} \right|^2 \right)}{2}$ and zero mean value. Based on Eq. (16), $Z_{1}^{3}$ follows the generalized central chi-square distribution with $2\left( D-2 \right) N_a$ degree of freedom and has the MGF as
\begin{small}
	\begin{align}
		M_{Z_{1}^{3}}\left( t \right) =\left[ \frac{1}{1-tN\left( \left| x_k \right|^2+\left| x_{\hat{k}} \right|^2 \right)} \right] ^\frac{N_l}{2},
	\end{align}
\end{small}where $N_l$ represents the number of same antennas in the selected AC and estimated AC.
Finally, the MGF of $Z_1$ can be obtained as
\begin{align}
	M_{Z_1}\left( t \right) =M_{Z_{1}^{1}+Z_{1}^{2}}\left( t \mid \boldsymbol{\mu }_1,\boldsymbol{\sigma }_{1}^{2} \right) M_{Z_{1}^{3}}\left( t \right) .
\end{align}
And the unconditional PEP of case 1 can be given by substituting Eq. (26) into (13) as
\begin{small}
	\begin{align}
		\mathrm{P_r}_{Z_1}\left( \left\{ r,k \right\} \rightarrow \{\hat{r},\hat{k}\} \right) =\frac{1}{\pi}\int_0^{\frac{\pi}{2}}{M_{Z_{1}^{1}+Z_{1}^{2}}\left( t\mid \boldsymbol{\mu }_1,\boldsymbol{\sigma }_{1}^{2} \right) M_{Z_{1}^{3}}\left( t \right) d\tau}.
	\end{align}
\end{small}
\par
\noindent 2) Second case $Z_2$: $\left\{ Z \mid r=\hat{r} \right\} $:
\par
In this case, we also can divide $Z_2$ into $Z_{2}^{1}+Z_{2}^{2}$ with
\begin{align}
	Z_{2}^{1}=&\left| x_k-x_{\hat{k}} \right|^2\sum\nolimits_{l\in V_r}^{}{\left| \sum_{q=1,q\ne l}^{N_a}{\sum_{i=\left( q-1 \right) N_E+1}^{qN_E}{\beta _{i,l}}}\alpha _ie^{\jmath\Psi _{i,l}} \right|^2} \nonumber
	\\
	&+\left| x_k-x_{\hat{k}} \right|^2\sum\nolimits_{l\in V_r}^{}{\left| \sum_{i=\left( l-1 \right) N_E+1}^{zN_E}{\beta _{i,l}}\alpha _i \right|}^2,
\end{align}
\begin{align}
	Z_{2}^{2}=&\left| x_k-x_{\hat{k}} \right|^2\times \nonumber
	\\
	&\sum_{n=1,n\ne l}^{N_r}{\left| \sum_{q=1}^{N_a}{\sum_{i=\left( q-1 \right) N_E+1}^{qN_E}{\beta _{i,n}\alpha _ie^{\jmath\left( \varphi _i-\omega _{i,l}-\theta _i \right)}}} \right|^2},
\end{align}
where $Z_{2}^{1}$ and $Z_{2}^{2}$ stand for the situations that $m=r$ and $m\ne r$, respectively. 
Firstly, by approximating $\beta _{i,l}\alpha _i$ as a Gaussian random variable with CLT, the mean value and variance of $Z_{2}^{1}$ can be obtained as $\mu _{Z_{2}^{1}}=\frac{N_aN_E\pi \left| x_k-x_{\hat{k}} \right|}{4}$ and $\sigma _{Z_{2}^{1}}^{2}=\frac{\left| x_k-x_{\hat{k}} \right|^2N_aN_E\left( 32-\pi ^2 \right)}{16}$. 
Then, by substituting $\mu _{Z_{2}^{1}}$ and $\sigma _{Z_{2}^{1}}^{2}$ into Eq. (23), the MGF of $Z_{2}^{1}$ can be obtained as
\begin{align}
	M_{Z_{2}^{1}}\left( t\mid \mu _{Z_{2}^{1}},\sigma _{Z_{2}^{1}}^{2} \right) =\left( 1-2\sigma _{Z_{2}^{1}}^{2}t \right) ^{-\frac{1}{2}}e^{\left( \frac{{t\mu _{Z_{2}^{1}}}^2}{1-2\sigma _{Z_{2}^{1}}^{2}t} \right)}.
\end{align}
Secondly, similar to the steps in case 1, $\Gamma _2$ can be approximately considered as a Gaussian random variable by applying CLT, which is expressed as
\begin{align}
	\Gamma _2=\sum_{q=1}^{N_a}{\sum_{i=\left( q-1 \right) N_E+1}^{qN_E}{\beta _{i,n}\alpha _ie^{\jmath\varPsi _{i,l}}x_k}},
\end{align}
with zero mean value and variance $\sigma _{\Gamma _2}^{2}=N$. Thus, $Z_{2}^{2}$ follows the generalized non-central chi-square distribution with $(N_r-N_a)$ degree of freedom and has the variance denoted as $\sigma _{Z_{2}^{2}}^{2}=\frac{NN_a\left| x_k-x_{\hat{k}} \right|^2}{2}$. Giving $\sigma _{Z_{2}^{2}}^{2}$ and following Eq. (23), the MGF of $Z_{2}^{2}$ can be simply  yielded.
Thus, the MGF of $Z_2$ can be obtained by multiplying MGF of $Z_{2}^{1}$ and $Z_{2}^{2}$, which is specifically given as $M_{Z_2}\left( t \right) =M_{Z_{2}^{1}}\left( t \right) M_{Z_{2}^{2}}\left( t \right) $,
and can be substituted into Eq. (13) to further obtain the unconditional PEP of case 2
$\footnote{Following the upper bound of Q-function, the PEP can be expressed as $\mathrm{P}\left( \left\{ r,k \right\} \rightarrow \{\hat{r},\hat{k}\} \right) \le \frac{1}{6}M_Z\left( -\frac{1}{N_0} \right) +\frac{1}{12}M_Z\left( -\frac{1}{2N_0} \right) +\frac{1}{4}M_Z\left( -\frac{1}{4N_0} \right) $. The closed-form expression can be obtained in this way, which is not extended in this paper due to the space limitation.}$
as
\begin{align}
	\mathrm{P}_{Z_2}\left( \left\{ r,k \right\} \rightarrow \{\hat{r},\hat{k}\} \right) =\frac{1}{\pi}\int_0^{\frac{\pi}{2}}{M_{Z_2}\left( t \right) d\tau}. 
\end{align}
Eventually, the union bound of ABER for the RASM scheme can be derived from the values of unconditional PEP by Eq. (13), which is given as
\begin{small}
	\begin{align}
			&P_{b}^{RASM}\le \frac{1}{MD}\times \nonumber
			\\&\sum_r{\sum_{\hat{r}}{\sum_k{\sum_{\hat{k}}{\frac{\mathrm{P}\left( \left\{ r,k \right\} \rightarrow \left\{ \hat{r},\hat{k} \right\} \right) e\left( \left\{ r,k \right\} \rightarrow \left\{ \hat{r},\hat{k} \right\} \right)}{\log _2\!\:\left( MD \right)}}}}},
		\end{align}
\end{small}where $e\left( \left\{ r,k \right\} \rightarrow \left\{ \hat{r},\hat{k} \right\} \right) $ represents the number of bits in error for the corresponding pairwise error event.
	Based on this analytical ABER, more insights of the BER performance can be found with the simulation results in Sec. \uppercase\expandafter{\romannumeral4}.
	
	\begin{figure}
		\centering
		\includegraphics[width=7.5cm,height=5.5cm]{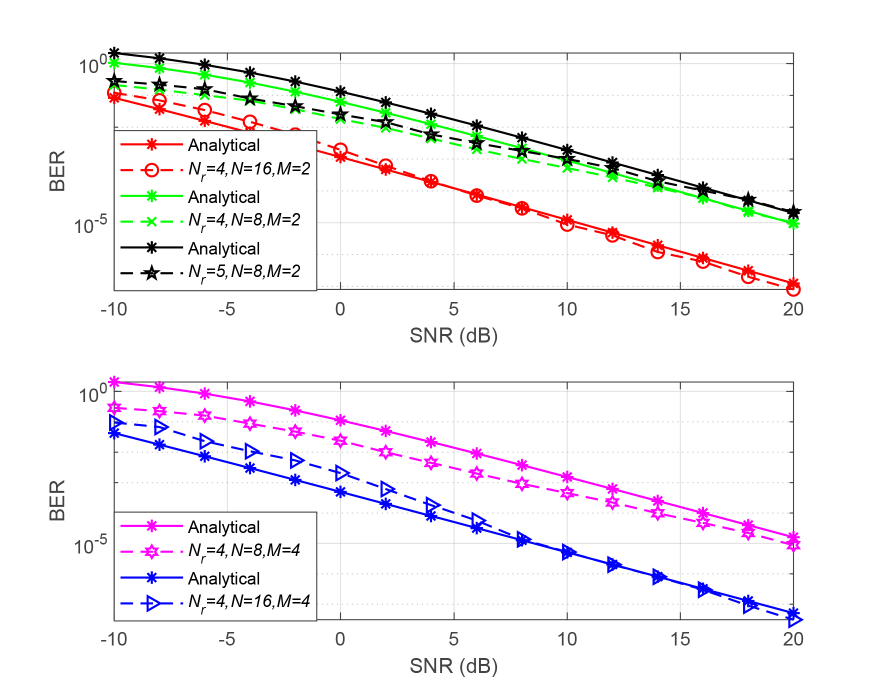}\\
		\caption{Analytical results and simulation ones in BER performance of the RASM scheme.}
	\end{figure}
	
	\begin{figure}
		\centering
		\includegraphics[width=7.5cm,height=5.5cm]{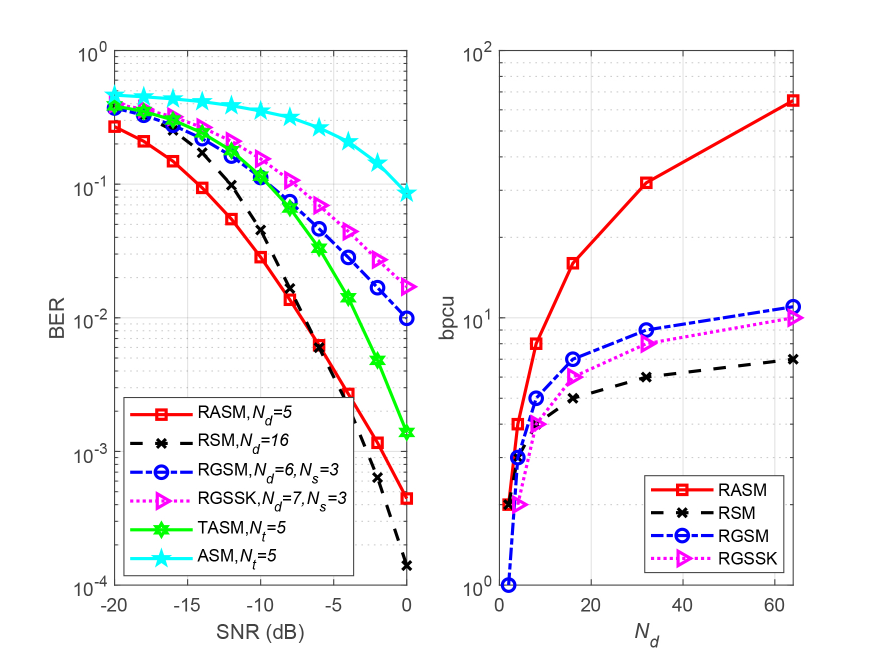}\\
		\caption{Comparison of BER performance and bpcu in the RASM, RSM, RGSM, and RGSSK schemes with $N=8$, $N_r=16$, and $M=2$.}
	\end{figure}
	
	\vspace{-7pt}
	
	\section{Simulation Results}
	The analytical and simulation results of the RASM scheme are presented in this section.
	Monte Carlo simulations are applied in the simulations, and all of experiments run for $1\times 10^6$ channel realizations.
	Similar to traditional schemes, $\frac{E_b}{N_0}$ is considered as the SNR, where $E_b=\left| x_k \right|^2=1$ represents the symbol's energy \cite{ref9}-\cite{ref14}.
	\par
	Fig. 2 shows the analytical and simulation results of BER with $N_r=4, 5$, $N=8, 16$, and $M=2, 4$, respectively.
	In Fig. 2, we can find that the analytical results, as theoretical upper bound, is approached to the simulation ones, especially when $N_r$ decreases, $N$  and SNR increase.
	Furthermore, the RASM scheme has lower than $10^{-3}$ BER when SNR$\gg 0$ dB with bpcu$=4$ and $N=16$, 
	which indicates the proposed one has satisfactory error performance under the lower SNR condition.
	Besides, the BER performance gets better with the arise of $N$, but it declines with the increase of $N_r$ and $M$
	$\footnote{This situation is merely considered in the received IM schemes, which is because increase of $N_r$ easily results in a large number of potential ACs, leading to higher complexity and lower accuracy in detection.}$ .
	\par
	Since all receive antennas are required to receive signals in transmitted IM schemes and the BER performance gets better with the increase of $N_r$.
	For fair comparison in same bpcu,
	we only need $N_d$ receive antennas to implement the RASM-based antenna selection while $N_r=16$ is assumed in all received IM schemes of Fig. 3.
	Fairly, bpcu of all schemes in the left figure of Fig. 3 is 5, and $N=8$ is given in all schemes in Fig. 3.
	Also, $M=2$ is for the RASM, RSM, and RGSM schemes.
	Thus, the RASM scheme has $N_d=5$, the RSM scheme has $N_d=16$, the RGSM scheme has $N_d=6$ and $N_s=3$, and the RGSSK scheme has $N_d=7$ and $N_s=3$, 
	where $N_s$ represents the number of selected antennas in the RGSM and RGSSK schemes [8-9].
	In addition, by comparing the BER performance of the TASM and ASM scheme with random AC selection, the BER performance of the RASM scheme significantly outperform that of the TASM and ASM scheme.
	The reason lies in the fact that within the TASM scheme, the AC indices undergo dual-stage channel fading, contrasting with the single stage in the RASM scheme, while additionally lacking the support provided by the RIS in the ASM scheme.
	According to Fig. 3, we find that the BER performance of the RASM scheme is better than that of the RGSM and RGSSK schemes.
	Besides, the BER performance of the RASM scheme is worse than that of the RSM scheme when SNR is larger than around $-2$ dB but better than RSM when SNR is lower than about -2 dB.
	Additionally, with the increase of $N_d$, the bpcu of the RASM scheme has significant improvement and is tremendously higher than other schemes especially with the high $N_d$,
	which further indicates the proposed scheme can provide higher SE in wireless transmission.
	\vspace{-7pt}
	\section{Conclusion}
	A novel RASM scheme is proposed in this paper as a trade-off selection applied in wireless transmission, maintaining satisfactory BER performance. 
	Additionally, the proposed scheme can further decrease the number of transmit antennas, reducing the complexity of hardware design at the transmitter. 
	The RASM scheme also provides higher SE by flexibly selecting more than one receive antenna in each time slot, meeting the requirements of high data rates in future communications.


\vspace{-7pt}
	\begin{thebibliography}{99}
		\bibitem{ref1} E. Basar et al., "Reconfigurable intelligent surfaces for 6G: Emerging hardware architectures applications and open challenges," \textit{IEEE Veh. Technol. Mag.}, vol. 19, no. 3, pp. 27-47, Sept. 2024.
		\bibitem{ref2}  M. Yue, Y. Peng, L. Xiong, C. Zhang, F. Al-Hazemi, and, M. M. Mirza, "Adaptive space shift keying for RIS-aided communication," \textit{IEICE Trans. Fundam. Electron. Commun. Comput.}, vol. E107-a, no.11, pp 1658-1662, Dec. 2024.
		\bibitem{ref3}  P. Yang, Y. Xiao, Y. Yu and S. Li, "Adaptive spatial modulation for wireless MIMO transmission systems," \textsl{IEEE Commun. Lett.}, vol. 15, no. 6, pp. 602-604, Jun. 2011.
		\bibitem{ref4}  K. Wang, C. -T. Lam and B. K. Ng, "How long can RIS work effectively: An electronic reliability perspective," in \textsl{Proc. IEEE 98th Veh. Technol. Conf. (VTC-Fall)}, Oct. 2023, pp. 1-6.
		\bibitem{ref5}  Y. Liu, C. Zhang, B. K. Ng, and C.-T. Lam, "Complex-valued neural network detection for RIS-assisted generalized spatial modulation," in \textsl{Proc. IEEE 100th Veh. Technol. Conf. (VTC-Fall)}, Oct. 2024, pp. 1-7.
		\bibitem{ref6}  C. Zhang, Y. Liu, B. K. Ng, and C.-T. Lam, "RIS-assisted differential transmitted spatial modulation design," \textit{Signal Process.}, vol. 230, p. 109767, 2025.
		\bibitem{ref7} E. Basar, "Reconfigurable intelligent surface-based index modulation: A new beyond MIMO paradigm for 6G," \textsl{IEEE T. Commun.}, vol. 68, no. 5, pp. 3187-3196, May 2020.	
		\bibitem{ref8}  Q. Jin, Y. Peng, F. Al-Hazemi, and J. Lee, "A pattern modulation based IRS scheme for space shift keying communication system," \textit{IEEE Commun. Lett.}, vol. 28, no. 6, pp. 1417-1421, 2024.
		\bibitem{ref9}  C. Zhang, Y. Peng, J. Li, and F. Tong, "An IRS-aided GSSK scheme for wireless communication system," \textsl {IEEE Comm. Lett.}, vol. 26, no. 6, pp. 1398-1402, Jun. 2022.
		\bibitem{ref10} C. Zhang and Y. Peng, "Received antenna array design of GSSK-based antennas selection for RIS-assisted communication," \textsl{IEEE Syst. J.}, vol. 17, no. 2, pp. 3366-3369, Jun. 2023.	
		\bibitem{ref11} T. Ma, Y. Xiao, X. Lei, P. Yang, X. Lei and O. A. Dobre, "Large intelligent surface assisted wireless communications with spatial modulation and antenna selection," \textsl{IEEE J. Sel. Areas Comm.}, vol. 38, no. 11, pp. 2562-2574, Nov. 2020.
		\bibitem{ref12} B. K. Ng and C. -T. Lam, "Characterization and optimization of coding performance in downlink NOMA with finite-alphabet inputs and finite blocklength," \textsl{IEEE Trans. Wireless Commun.}, vol. 23, no. 4, pp. 2796-2811, Apr. 2024.
		\bibitem{ref13}  H. Xu, C. Zhang, Q. Wu, B. K. Ng, C. -T. Lam and H. Yanikomeroglu, "FTN-assisted SWIPT-NOMA design for IoT wireless networks: A paradigm in wireless efficiency and energy utilization," \textit{IEEE Sensors J.}, vol. 25, no. 4, pp. 7431-7444, 15 Feb.15, 2025.
		\bibitem{ref14} Q. Jin, Y. Peng, F. Al-Hazemi, and J. Lee, "A pattern modulation based IRS scheme for space shift keying communication system," \textit{IEEE Commun. Lett.}, vol. 28, no. 6, pp. 1417-1421, Oct., 2024.
	\end{thebibliography}
\end{document}